\documentclass[twocolumn,showpacs,preprintnumbers,amsmath,amssymb]{revtex4}

\usepackage{graphicx}
\usepackage{dcolumn}
\usepackage{bm}
\usepackage{subfigure}

\begin{document}
\title{Intrinsic Defects and Electronic Conductivity of TaON: First-Principles Insights}

\author{Shiyou Chen and Lin-Wang Wang}
\affiliation{Materials Sciences Division, Lawrence Berkeley National Laboratory,
One Cyclotron Road, Mail Stop 66, Berkeley, CA 94720}
\date{\today}

\begin{abstract}
As a compound in between the tantalum oxide and nitride, the tantalum oxynitride TaON is expected to combine their advantages and act as an efficient visible-light-driven
photocatalyst. In this letter, using hybrid functional calculations we show that TaON
has different defect properties from the binary tantalum oxide and nitride: (i) instead of
O or N vacancies or Ta interstitials, the $O_N$ antisite is the dominant defect, which determines its intrinsic n-type conductivity and
the p-type doping difficulty; (ii) the $O_N$ antisite has a shallower donor level than O or N vacancies, with a delocalized distribution
composed mainly of the Ta $5d$ orbitals, which gives rise to better electronic conductivity in the oxynitride than in the oxide and nitride.
The phase stability analysis reveals that the easy oxidation of TaON is inevitable under O rich conditions, and a relatively O poor
condition is required to synthesize stoichiometric TaON samples.

\end{abstract}

\maketitle
The splitting of water into H$_2$ and O$_2$ using semiconductor
photocatalysts has been considered as a clean and
renewable way to utilize solar energy. One key issue for realizing this application is to search for suitable photoelectrode semiconductors
which can absorb visible light.\cite{gai-036402-2009,walter-6446-2010,majumder-053105-2011} Owing to their good chemical stability in aqueous
solution, transition metal oxides are the subject of most previous
studies.\cite{walter-6446-2010,maeda-7851-2007}  However, transition metal oxides usually have large band gaps ($\geq$ 3 eV),
limiting the absorption of visible light and setting an upper limit
on the energy conversion efficiency at $\sim$2$\%$.\cite{maeda-7851-2007,walter-6446-2010} In contrast,
the transition metal nitrides have smaller band gaps due to
N 2p orbitals being shallower than O 2p orbitals, but the easy oxidation of nitrides
in aqueous solution makes them degraded quickly. Being in between the oxides and nitrides, the oxynitrides are believed to have intermediate
properties, such as smaller band gaps than oxides and better
stability than the nitrides.\cite{maeda-7851-2007,walter-6446-2010,wang-065501-2010} A well known example, tantalum oxynitride
TaON, which has a band gap around 2.5 eV, between those of
Ta$_2$O$_5$ (3.9 eV) and Ta$_3$N$_5$ (2.1 eV), was found to have high
quantum yield for nonsacrificial visible-light-driven water
splitting.\cite{maeda-5868-2010,maeda-3057-2011,abe-11828-2010} Furthermore, TaON has appropriate alignment of its valence and
conduction band edge relative to the OH$^-$/ O$_2$ oxidation and H$^+$/ H$_2$ reduction potentials, make it possible to produce H$_2$ and
O$_2$ from water even without an externally applied
bias.\cite{maeda-7851-2007,abe-11828-2010}

Despite the experimental progress in the photocatalytic performance of TaON, it is currently not well
understood what intrinsic defects contribute to the observed n-type conductivity.\cite{maeda-5868-2010,nakamura-8920-2005}
For transition metal oxides and nitrides, usually the anion (O or N)
vacancies and cation interstitials play an important role in determining its intrinsic n-type conductivity,\cite{janotti-085212-2010,agoston-245501-2009} thus it is natural to ask
if the same situation exists in TaON. Previously the defects related to the anion vacancies or reduced Ta were taken as the origin of the observed n-type character,\cite{maeda-5868-2010} but there
was no detailed study about whether these defects are easy to form energetically. On the other hand, since the Z-scheme water splitting devices require n-type semiconductors
as the photoanode for O$_2$ evolution and p-type semiconductors as the photocathode for H$_2$ evolution,\cite{walter-6446-2010,maeda-5868-2010,maeda-3057-2011}
the ability to dope TaON to n-type or p-type becomes important. As far as we know, most experiments observed only n-type
conductivity in TaON, which is one of the reasons for its better
 activity in O$_2$ evolution than H$_2$ evolution,\cite{maeda-7851-2007,abe-11828-2010}  and it is not clear if p-type doping is possible. In this letter, we try to use the density functional theory calculation to shine
a light on the defect properties of TaON and their influence on the electronic conductivity.

Using the generalized gradient approximated (GGA) exchange-correlation functional, previous calculations have
shown that the ground state $\beta$-TaON structure has an indirect band gap at about 1.8 eV,\cite{schilling-2931-2007,lumey-2173-2003,fang-1248-2001,li-6917-2010,yashima-588-2007}
smaller than the experimental value 2.5 eV,\cite{maeda-7851-2007} which is common
according to the well-known band gap underestimation by GGA. To avoid such band gap error and its significant influences on the defect formation energy and transition energy level,\cite{du-115217-2009, agoston-245501-2009, clark-115311-2010}
in our calculation we used
the non-local hybrid functional (HSE06\cite{hse06})
in which 16\% ($\alpha$) of the semi-local GGA exchange potential is replaced by
screened Fock exchange.  Here the ratio $\alpha$=16\% is used because the calculated band gaps
agree better with experiments for such kind of oxides which has its top valence band mainly composed of the O $2p$ orbitals and its bottom conduction band mainly composed of
the cation $d$ orbitals.\cite{bilc-165107-2008} Our test also shows that the calculated band gap using $\alpha$=16\% gives an indirect band gap of 2.61 eV,
closer to the experimental value 2.5 eV,\cite{maeda-7851-2007} than that (3.02 eV) calculated using the standard $\alpha$ value 25\%. A small
$\alpha$ value 20\% had also been used in the study of defects in TiO$_2$.\cite{janotti-085212-2010} Our spin-polarized calculation is performed using the VASP
code,\cite{vasp1} with the frozen-core projector augmented-wave potentials\cite{paw}, an energy cutoff of 400 eV for plane wave basis
and single \textit{k}-point for a 96-atom supercell.

\begin{figure*}
\scalebox{0.8}{\includegraphics{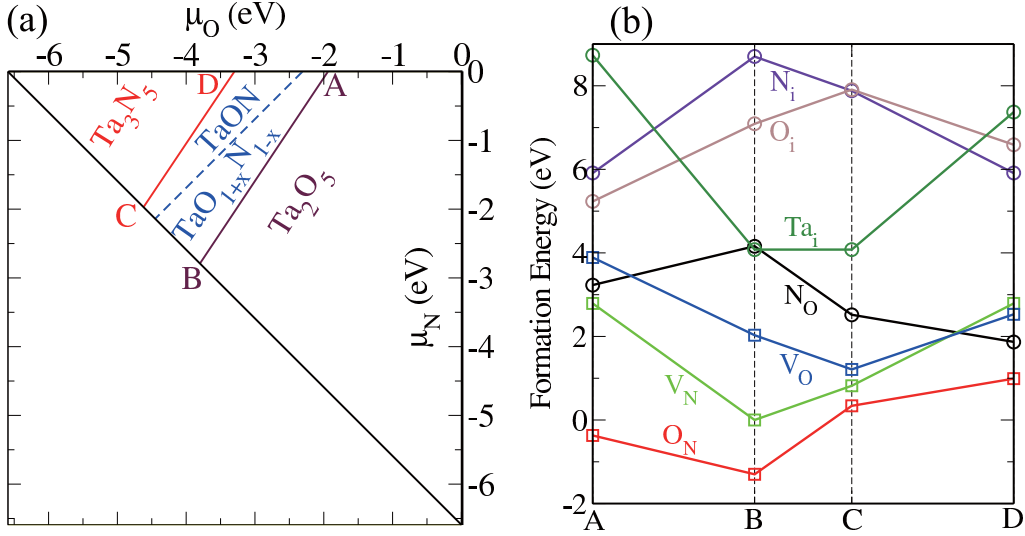}}
\caption{\label{fig:epsart1} (Color online) (a) The calculated chemical potential
range of O and N that stabilizes TaON, (b) the formation energy
change as a function of the chemical potential points A, B, C and D
as shown in (a) for neutral defects.}
\end{figure*}

The supercell approach is used to calculate the defect formation
energy $\Delta H(\alpha,q)$, i.e., a
defect $\alpha$ in the charge state $q$ is placed in a 96-atom
TaON supercell. From the total energy $E(\alpha,q)$ of
the supercell with a defect and that of the perfect supercell
$E(host)$, $\Delta H(\alpha,q)$ is calculated according
to
\begin{equation}\label{e-1}
\begin{split}
\Delta H(\alpha,q) = & E(\alpha,q)-E(host)+\sum_i n_i(E_i+\mu_i)\\
& +q[\epsilon_{VBM}(host)+E_F],
\end{split}
\end{equation}
where $\mu_i$ is the elemental chemical potential of constituent $i$ ($i$=Ta, O, N)
referenced to the total energy $E_i$ of its pure elemental phase (Tantalum metal bulk, isolated O$_2$ and
N$_2$ molecules). $\mu_i$=0
represents the limit where the element is so rich that their pure
phase can form. $E_F$ is the Fermi energy level referenced to
$\epsilon_{VBM}(host)$, the valence band maximum (VBM) eigenenergy of the host TaON. $n_i$ is
the number of atom $i$, and $q$ is the number of electrons exchanged
between the supercell and the corresponding thermodynamic reservoir
in forming the defect. For $\Delta H(\alpha,q)$ of charged defects,
the first-order Makov-Payne correction is taken for the
spurious electrostatic interaction caused by the limited size of the supercell. More details about the
calculation methods can be found in Ref. \cite{wei-337-2004,lany-235104-2008,chen-245204-2010}

As we can see in Eq. (1), the defect formation energy depends on the elemental chemical potential $\mu_i$ of O, N and Ta,
thus we will first determine the chemical potential range. To avoid the formation of pure elemental phase,
$\mu_{Ta}$ $<$ 0, $\mu_{O}$ $<$ 0, $\mu_{N}$ $<$ 0 must be satisfied;
to avoid the formation of binary compounds Ta$_2$O$_5$ and Ta$_3$N$_5$,

\begin{equation}\label{e-2}
\begin{split}
&2\mu _\mathrm{Ta}+5\mu _\mathrm{O} < \Delta \mathrm{H_f(Ta_2O_5)}=-18.14~\mathrm{eV}\\
&3\mu _\mathrm{Ta}+5\mu _\mathrm{N} < \Delta \mathrm{H_f(Ta_3N_5)}=-9.45~\mathrm{eV}\\
\end{split}
\end{equation}
should be satisfied, where $\Delta \mathrm{H_f}$ stands for the calculated formation energy of $\delta$-Ta$_2$O$_5$ and Ta$_3$N$_5$.\cite{sahu-165202-2004,fang-1248-2001}
Considering that TaON should be thermodynamically stable, $\mu _\mathrm{Ta}+\mu _\mathrm{O}+\mu _\mathrm{N} = \Delta \mathrm{H_f(TaON)}=-6.23~\mathrm{eV}$ should also be kept.
Following all these requirements, the chemical potential that stabilizes TaON is limited in the region surrounded by A, B, C and D points shown in Fig. 1(a).
As we can see, Ta$_2$O$_5$ phases will be formed if $\mu_{O}$ is too high (O rich), while Ta$_3$N$_5$ phases will be formed if $\mu_{O}$ is too low (O poor), showing the fact that the richness of
O is an important factor influencing the TaON stability.

\begin{figure*}
\scalebox{1.0}{\includegraphics{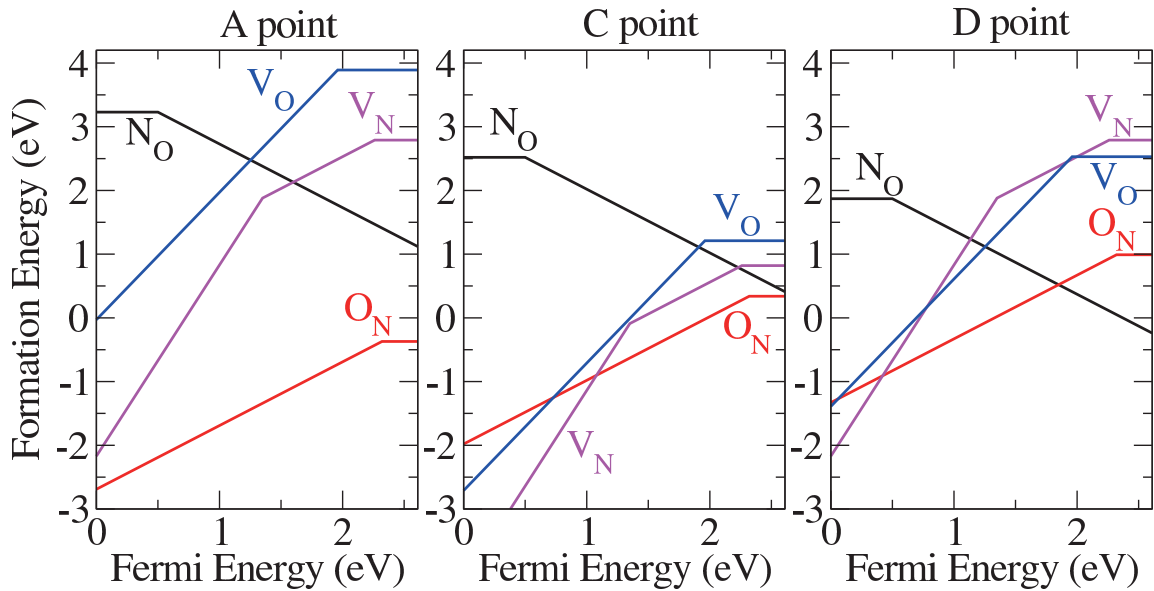}}
\caption{\label{fig:epsart1}(Color online) The change of the defect formation
energy as a function of the Fermi energy at the chemical potential
points A (left), C (center) and D (right). The slope of the change
shows the charge state q, and only the most stable charge state is
plotted for a certain Fermi energy. The turning points show the
transition energy levels at which the charge states change.}
\end{figure*}

Under the chemical potential condition that stabilizes TaON, we further calculated the formation energy of a series of intrinsic defects, including the O vacancy $V_{O}$, N vacancy
$V_{N}$, $O_N$ antisite with one N atom replaced by O, $N_O$ antisite, O interstitial $O_i$, N interstitial
$N_i$, as well as the Ta vacancy $V_{Ta}$ and Ta interstitial $Ta_i$. The results for the neutral defects are plotted in Fig. 1(b). Obviously the neutral $O_N$ antisite has lower
formation energy than all other intrinsic defects, independent of the chemical potential. The two anion vacancies ($V_{O}$ and $V_{N}$) and $N_O$ antisite have relatively higher
energy, and all interstitials have much higher formation energy (the results of $V_{Ta}$ is not plotted due to a formation energy higher than 14 eV). This order in the formation energy
can be understood considering the smaller chemical and size change caused by the substitution of a N atom by an O atom, than the substitution of a N (or O) atom by a vacancy.
Compared with the situation in the binary Ta$_2$O$_5$ and Ta$_3$N$_5$ compounds where the anion vacancies or cation interstitials are the dominant defects,
in ternary TaON the anion antisites become the dominant defects, which may have a different contribution to the electronic conductivity.

One point that should be mentioned is that the $O_N$ antisite has a negative formation energy at the chemical potential conditions A and B, i.e., the defects will be formed spontaneously,
which will cause a sample to be non-stoichiometric,
like TaO$_{1+x}$N$_{1-x}$ with x$>$0. The easy formation of $O_N$ antisites explains the experimental observation that
the TaON surface is oxidized
when the photocatalytic reactions begins and the O$_2$ gas is evolved (equivalent to increasing $\mu_{O}$, O rich).\cite{maeda-7851-2007}
In Fig. 1(a) we show that, the stoichiometric TaON is stabilized on the left side of the dashed line, while the non-stoichiometric TaO$_{1+x}$N$_{1-x}$ is stabilized on the right side
due to the abundant formation of $O_N$ antisites. According to this analysis, to synthesize the single-phase and stoichiometric TaON samples,
relatively O poor growth conditions should be used, like at C and D points where all intrinsic defects have positive formation energy.

Now we will analyze the influence of intrinsic defects on the electronic conductivity. In Fig. 2 we plot the formation energy change of different charged defects as the Fermi energy
level $E_F$ shifts from the valence band maximum (VBM, $E_F$=0 eV) to conduction band minimum (CBM, $E_F$=2.61 eV), under the three different chemical potential condition A, C and D.
According to Equ. (1) the formation energy increases when the Fermi energy increases for positively charged donors, such as $O_N$ antisite and $V_{O}$ and $V_{N}$ vacancies, while
decreases for negatively charged acceptors such as $N_O$ antisite. Under conditions A and C, the donor defect $O_N$ always has a lower formation energy than the lowest energy acceptor
defect $N_O$, determining the intrinsic n-type conductivity. Under the condition D, $O_N$ has a lower formation energy than $N_O$ when the Fermi energy level is closer to the VBM, but
the order is reversed as the Fermi energy approaches the CBM, i.e., the negatively charged (q=-1) $N_O$ acceptor can form spontaneously, killing the electron carriers and pinning the
Fermi energy at about 0.8 eV below the CBM. This further indicates that although n-type conductivity can be observed due to the formation of $O_N$ under this condition, the density of the free electrons is limited.

From the above discussion, it is now clear that the dominant $O_N$ donor defects determine the intrinsic n-type conductivity of TaON.
On the other side, because the formation energy of $O_N$ decreases to negative when the Fermi energy shifts down to near VBM, it will kill the hole carriers and pin the Fermi
energy closer to the CBM if we want to dope TaON to p-type, resulting in the p-type doping difficulty. This fact agrees with the experimental
observation that only n-type conductivity has been observed for TaON samples up to now, and also indicates that TaON is better to be used as a
photoanode materials in the Z-scheme water-splitting system.

\begin{figure*}
\scalebox{1.0}{\includegraphics{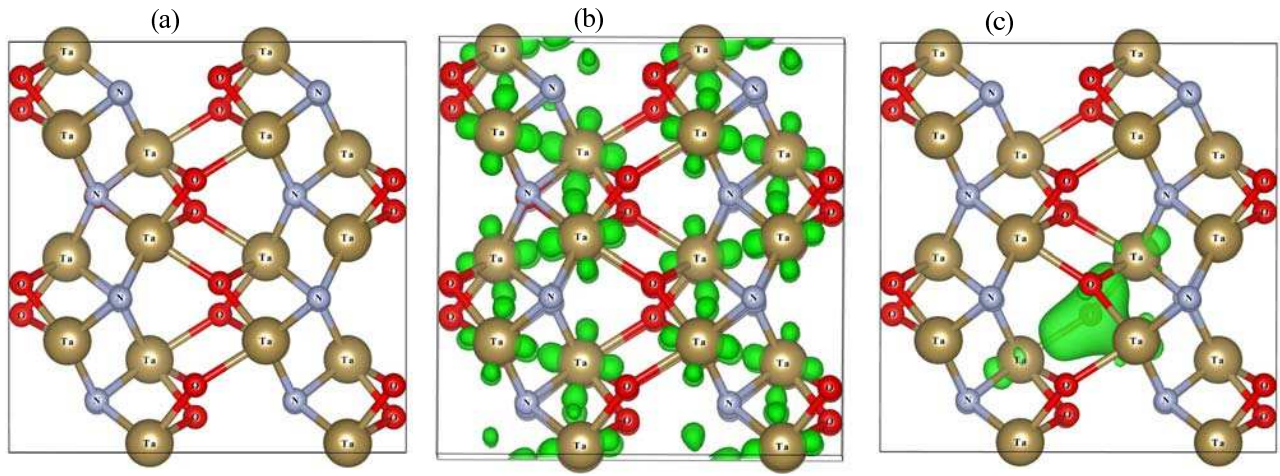}}
\caption {\label{fig:epsart1}
(Color online) The plot of the TaON crystal structure
(a), the charge density contour (in green) of the single-particle
$O_N$ (b) and $V_{O}$ (c) defect states. The circles in oliver, red
and steelblue show the Ta,O and N atoms respectively, and the dashed
circles show the defect sites.}
\end{figure*}

The turning points in Fig. 2 stand for the transition energy levels at which the charge states of defects change, eg. the $\epsilon(0/+)$ level of $O_N$ is 0.29 eV below the CBM,
which is much shallower than the $\epsilon(0/2+)$ of $V_{O}$ (0.5 eV below CBM) and the $\epsilon(1+/3+)$ of $V_{N}$ (0.65 eV below CBM). Since the deep defect levels in the band gap
usually act as the recombination center of electron-hole pair, and give bad electronic conductivity, the shallower transition energy level
associated with the dominant $O_N$ donor defect indicates that TaON is a better photoelectrode material with fewer recombination center and better electronic conductivity than
the binary Ta$_2$O$_5$ and Ta$_3$N$_5$, in which the dominant anion vacancies or cation interstitials produce deep levels in the band gap.\cite{shin-116108-2008} It should be noted that, the absolute position (0.29 eV below CBM) of
$O_N$ $\epsilon(0/+)$ donor level is not so shallow, but the contour plot of charge density in Fig. 3(b) shows that the $O_N$ donor state is very delocalized over the whole supercell,
composed mainly of the Ta $5d$ orbitals, which shows directly that this donor state will give good electronic conductivity. In contrast, the $V_{O}$ donor state is much more localized
as shown in Fig. 3(c) with distribution mainly around the vacancy, despite its transition energy level is just 0.36 eV deeper than $O_N$ $\epsilon(0/+)$.

In Fig. 2, when the Fermi energy is close to the CBM (n-type sample), the two lowest energy defects in TaON are $O_N$ donor and $N_O$ acceptor. Because of the electronic compensation and the electrostatic attraction
between them, the $O_N+N_O$ antisite pair (equal to the exchange of O and N atoms) may also have low formation energy and become popular in TaON, explaining the random distribution of O/N anions in the earlier
experiments.\cite{fang-1248-2001}
Since the $O_N$ donor level $\epsilon(0/+)$ is 0.29 eV below the CBM and the $N_O$ $\epsilon(-/0)$ acceptor level is
0.50 eV above the VBM, we may expect that the donor-acceptor recombination will generate a photoluminescence peak at around 2.61-0.29-0.50=1.82 eV, which is consistent with the recent experiment
where a peak at 690 nm is clearly observed,\cite{maeda-5868-2010} indicating that the $O_N$ and $N_O$ antisite defects are popular in the sample.

In conclusion, the phase stability, the formation of intrinsic defects and their influences on the electronic conductivity of TaON are studied based on the
density functional theory calculation. Stoichiometric TaON is found to be stable under a relatively O poor condition, and can be easily oxidized due to the easy formation
of $O_N$ antisites. Different from the situation in oxides and nitrides, the $O_N$ antisite becomes the dominant intrinsic defect in TaON, explaining the n-type conductivity
observed experimentally, and also determining the p-type doping difficulty in this system. In contrast with the deep donor states associated with
the O and N vacancies, the donor state of the $O_N$ antisite is shallow and has a delocallized distribution, which indicates a better electronic conductivity in TaON than in Ta$_2$O$_5$ and Ta$_3$N$_5$.

We thank Dr. Joel W. Ager, Cheng-Hao Wu, Sefa Dag, Shu-Zhi Wang and Ajay Yadav for their helpful discussion, and Chris Barrett for his modification of the manuscript.
This work is supported by the Joint Center of Artificial Photosynthesis and the BES/SC of the U.S. Department of Energy under the contract No. DE-AC02-05CH11231, and the computation
is performed using the NERSC and NCCS supercomputers.


\end{document}